\documentclass[conference,10pt]{IEEEtran}
\usepackage{epsfig}
\usepackage{graphicx}
\usepackage{float}
\usepackage{amsmath}

\begin{document}
\title{Load Repartition for Congestion Control in Multimedia Wireless Sensor Networks with Multipath Routing}

\author{\authorblockN{Moufida Maimour}\authorblockA{CRAN, Nancy
University, CNRS\\Moufida.Maimour@cran.uhp-nancy.fr}\and
\authorblockN{C. Pham}\authorblockA{LIUPPA, University of
Pau, France\\Congduc.Pham@univ-pau.fr}\and\authorblockN{Julien Amelot}\authorblockA{CRAN, Nancy
University, CNRS\\Julien.Amelot@esial.uhp-nancy.fr}}

\maketitle

\begin{abstract}
Wireless sensor networks hold a great potential in the deployment of
several applications of a paramount importance in our daily
life. Video sensors are able to improve a number of these applications
where new approaches adapted to both wireless sensor networks and
video transport specific characteristics are required. The aim of this
work is to provide the necessary bandwidth and to alleviate the
congestion problem to video streaming. In this paper, we investigate
various load repartition strategies for congestion control mechanism
on top of a multipath routing feature. Simulations are performed in
order to get insight into the performances of our proposals.
\end{abstract}
\IEEEpeerreviewmaketitle


\section{Introduction}\label{intro}

This paper addresses the problem of congestion control
in wireless sensor networks (WSN). Congestion control in WSN is
particularly difficult as a WSN can usually remain idle for a long
period of time and then suddenly become active in response to a
detected event, generating a large amount of
information from the sources (sensors) to a sink. Even if
an event is a few bytes long, the high number of events due to high
reporting rates will rapidly create shortage of resources (buffer
space, battery life) in the WSN leading to congestion and consequently
packet/event drops.

In addition to traditional sensing network infrastructures, a wide
range of emerging wireless sensor network applications for object
detection, surveillance, recognition, localization, and tracking can
be strengthened by introducing a visioning capability. Nowadays, such
applications are possible since low-power sensors equipped with a
visioning component \cite{Cyclops,Panoptes} already
exist. In these Wireless Multimedia Sensor Networks (WMSN) \cite{AKY07}, congestion control is of prime importance due to the inherently high rate of
injection of multimedia packets in the network (video traffic is in
the order of 250 kbit/s to 500 kbit/s).

The problem of congestion control have been addressed in many works
along with a transport layer protocol proposition. CODA \cite{WAN03},
ESRT \cite{SAN03}, RMST \cite{STA03} are some of
those propositions to name a few. We believe that cross-layer design
holds great potential for addressing video
transport challenges. This work is a first step in this direction by
addressing congestion control with a multi-path routing facility. In this paper, we investigate various load repartition strategies on top of a multipath
routing feature for congestion control. The target application is
video transmission and in order to keep the video quality unchanged,
we avoid decreasing the transmission rate. Instead, a video flow is
split on multiple paths if there are some available. There have been
many work on load balancing or repartition but none for the best of
our knowledge has specifically been addressed for or been used on wireless sensor networks. This paper tries to show whether load repartition is useful or not, and if so, where are the expected gains.

The work presented recently in \cite{POP06} is the closest to our
proposition since their End-to-end Packet Scatter (EPS) split traffic
on multiple paths, in an attempt to spread network load on a wider area based on the Biased Geographical Routing protocol. However, the complexity of BGR that requires location features and of the congestion control mechanism that adds an In-Network-Packet-Scatter prior to the EPS mechanism is much higher than our proposition. EPS also is much more costly in terms of control messages. We only split the flows at the source and need less control feedback messages. We think that simplicity is of prime importance and should serve as a guideline when designing congestion control on WMSN.

The load repartition strategies vary from the simplest one which
distributes uniformly the traffic on all available paths
simultaneously to more complex strategies with explicit congestion
notifications (CN) from congested nodes towards the sources. In these
cases, on reception of a CN, a source will try to balance its traffic
on available paths in order to keep its sending rate unchanged while
reducing the amount of data sent on the current active paths.
Congestion inferences could be based on the queue length at
intermediary nodes such as in CODA or ESRT. At this point, we must
state that the proposed solutions does not seek to obtain the optimal
load repartition on all existing paths, but rather to react as quickly
as possible to congestion to avoid packet losses in very
resource-constrained devices. Therefore we plaid for a simple
mechanism that limits both the number of exchanged control messages
and the complexity at the sources.

The proposed mechanisms can be used with any multipath routing layer
where explicit congestion notification, possibly implemented at a
higher layer than the network layer, are available from intermediary
nodes. However, in this paper, we study and propose the usage of SLiM
(Simple Lifetime-based Multipath routing protocol) that was previously
described in \cite{SLIM} and that provides multipath routing from a
set of sources to a given sink with a path's lifetime criterion.


The paper is organized as follows. Section \ref{networkModel} presents
the network model with the different assumptions considered in this
work. The SLiM multipath routing protocol is also briefly presented
for the purpose of making the paper self-reading. Section \ref{CC}
presents the various load repartition strategies for congestion
control on top of SLiM. Some simulation results are presented in
section \ref{simulation} before concluding.

\section{Network Model}\label{networkModel}

We consider a wireless sensor network with video sensors located in
strategic locations and other non visual sensors distributed randomly
in a field. A video sensor is asleep and is only waked up when alerted
by other non visual sensors upon target detection or as a response to
a request. In this paper we consider the case of multiple video
sensors (referred to as the sources) being reporting video information
to the sink at relatively the same time.

Video applications are considered as semi-reliable ones where some
losses are tolerated but a minimum data rate is required from the
beginning of the transmission. In order to be able to satisfy this
requirement, we investigate the use of multiple paths so a maximum
bandwidth can be supplied. Therefore, we assume that a multipath
routing protocol is available. In this paper, we use SLiM \cite{SLIM}
but any other multipath routing protocol able to build and
maintain at the same time more than one path can be used by our
congestion control scheme.

SLiM, with only local topology knowledge, provides to a source and all
intermediary nodes the knowledge of all available paths to the
sink. It adopts the sink-initiated approach 
where the sink is the originator of a request. The sink floods the
network with a request until the sensor, referred to as the source,
having the target in its field of view is reached. With one flooding,
multiple paths are built and maintained at intermediate nodes towards
the sink. In SLiM, a request is identified using a path id that
corresponds to the first crossed sensor's id from the sink to the
source. Paths are built with respect to a quality metric specified by
the application. This metric can be the path length, its available
energy, an estimation of its lifetime or any other metric depending on
the application requirements.

Each sensor is able to create, maintain and update a path table that
records the different paths to the sink. The table contains an entry for each path with the following fields :
\begin{itemize}
\item {\it pid}, the path id,
\item {\it inUse}, a flag, when set indicates that the
  corresponding path is currently in use,
\item {\it nextNode}, the next hop towards the sink on this path,
\item {\it quality}, an estimation of the associated quality metric
for this path.
\end{itemize}

\begin{figure}[H]
  \centerline{\psfig{figure=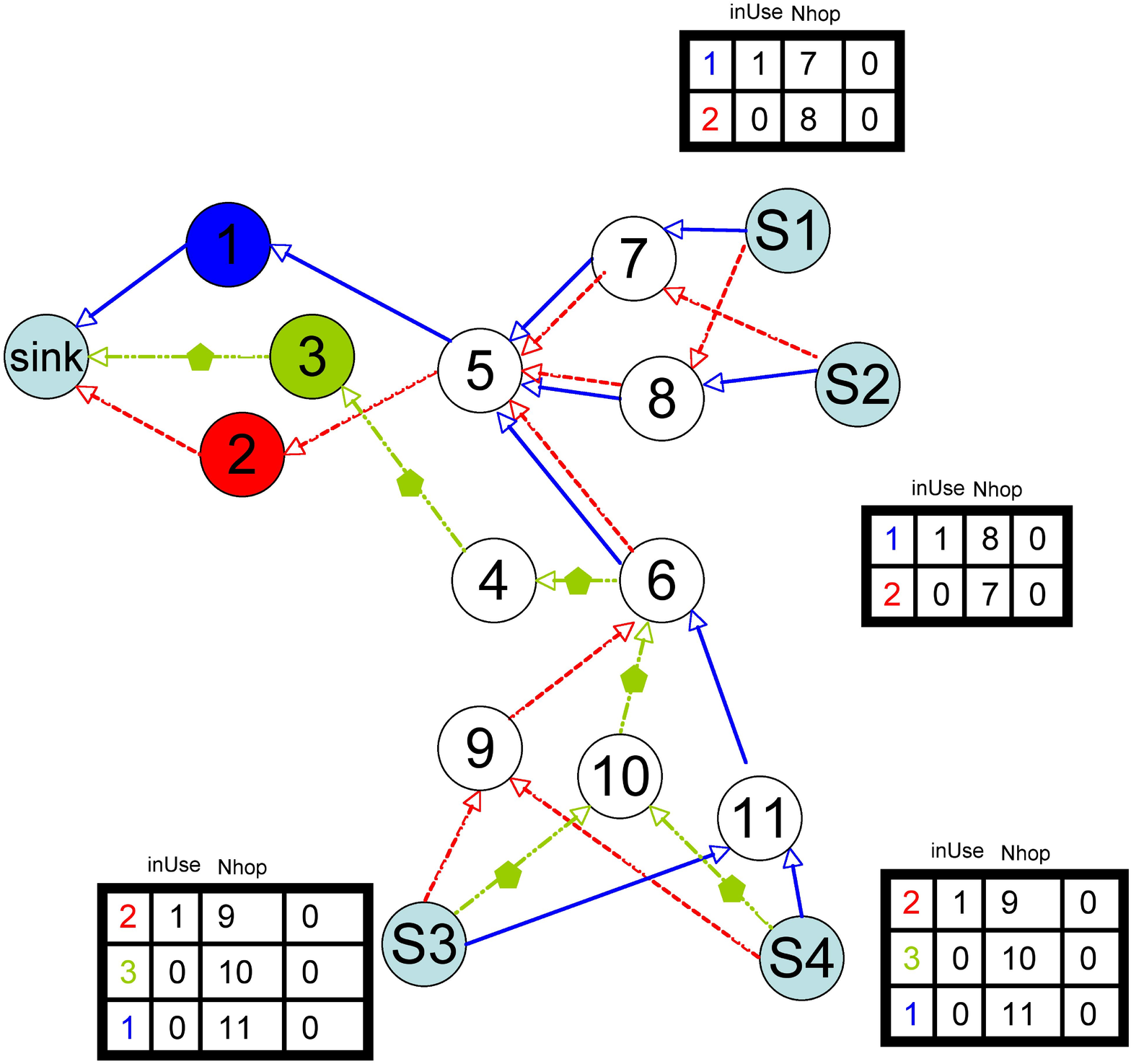,width=.8\linewidth}}
  \caption{Network model with multiple paths to the sink.}
  \label{MultiPath-1}
\end{figure}

Figure \ref{MultiPath-1} shows a configuration typically built
by SLiM. This scenario shows 15 sensor nodes and 1 sink. Among the 15
sensor nodes, there are 4 sources identified as S1, S2, S3 and S4. We
can see that there are 3 different paths with path id 1, 2 and 3,
named after the first crossed node's id from the sink. S1 and S2 have path
id 1 and 2 in their forwarding table. S3 and S4 have path id 1, 2, and
3 in their forwarding table. The right-most column keeps the rate
repartition as will be explained later on. Note that SLiM avoids
constructing different path id at a source with the same
predecessor. This was done in order to limit congestion when no
congestion control was defined on top of SLiM.

For what follows, the sink is never the bottleneck nor in term of
bandwidth nor in term of energy. It can have its battery
recharged or replaced in real applications. In contrast, all the sensors
have limited energy, are supposed to be stationary but with the ability to dynamically vary their transmission power. 

We assume that an addressing scheme is available. Globally unique
addresses can be very expensive in terms of bandwidth and power
consumption. Instead we consider a local addressing scheme as the one
proposed in \cite{Muneeb04}. We use an address for a sensor that can
be reused by an other one located sufficiently far away. We assume,
however, a uniquely assigned address to the sink in order to
distinguish it from the other sensors. 


\section{Load repartition for congestion control}\label{CC}

In this paper we investigate the use of load repartition mechanisms
for the purpose of congestion control. In figure \ref{MultiPath-1},
after SLiM has constructed the path configuration and forwarding
tables, we assume that S1 and S2 use path id 1 as the default path
whereas S3 and S4 use path id 2. This information is stored in the
source's forwarding table with the $inUse$ field.
An additional field in the forwarding table keeps the current data
rate (or an estimation if the exact data rate is not known) sent on
the path. We assume that each source stores paths in the order of
decreasing quality.
 
\subsection{Load repartition strategies}

We define 3 load repartition strategies for congestion control, from
mode 1 to mode 3. For the purpose of comparison, mode 0 refers to the
no load repartition scenario in which a source uses the same path (the
best path in term of lifetime with SLiM) without any congestion
control concerns. In mode 1 the source uses all the available paths to
a sink from the beginning of the transmission. The traffic is then
uniformly load-balanced on these paths. Mode 0 and mode 1 therefore represent the 2 end-points in the load repartition strategies design space.

In modes 2 and 3, explicit congestion notifications are used. At every
intermediary node, when the reception queue occupancy is greater than
a given threshold (80\% of total buffer space for instance) or when the collision rate is above a given threshold, a Congestion Notification (CN) message is sent back to the sources for
each path id known by the node. A CN message contains the node id and the path id: CN(nid, pid).
For simplicity, we assume that each source sends 1
data flow identified by the source id. A source $S_i$ should react to
a CN message if the path id contained in the CN message corresponds to
an active path in its local forwarding table. The basic principle
behind these load repartition strategies is to make each source aware
of a congestion on path $i$ and reacting to it by load-balancing the
current traffic on this path on a larger number of paths. Selected
paths at the source are then marked as active with the $inUse$ flag,
and the data rate repartition for each path is kept in the forwarding
table. In the following paragraphs, we will describe mode 2 and 3.

\begin{itemize}

\item {\bf Mode 2}. The source starts initially with one path. For
each CN(nid, pid) message received, the source adds a new path (the
first available path different from $pid$ that is non active)
until all available paths are marked as active. The load is uniformly
distributed on the number of active path. It is therefore an
incremental approach.

\item {\bf Mode 3}. The source starts initially with one path. Upon
reception of a CN(nid, pid) message the source will uniformly balance
the traffic {\bf of path $pid$} on all available paths (including path
$pid$). Therefore depending on the number of CNs received for each path, the transmission rate is not the same on all the active paths as opposed to mode 2.




\end{itemize}

\subsection{Detailed example of mode 3}

\begin{figure}
  \centerline{\psfig{figure=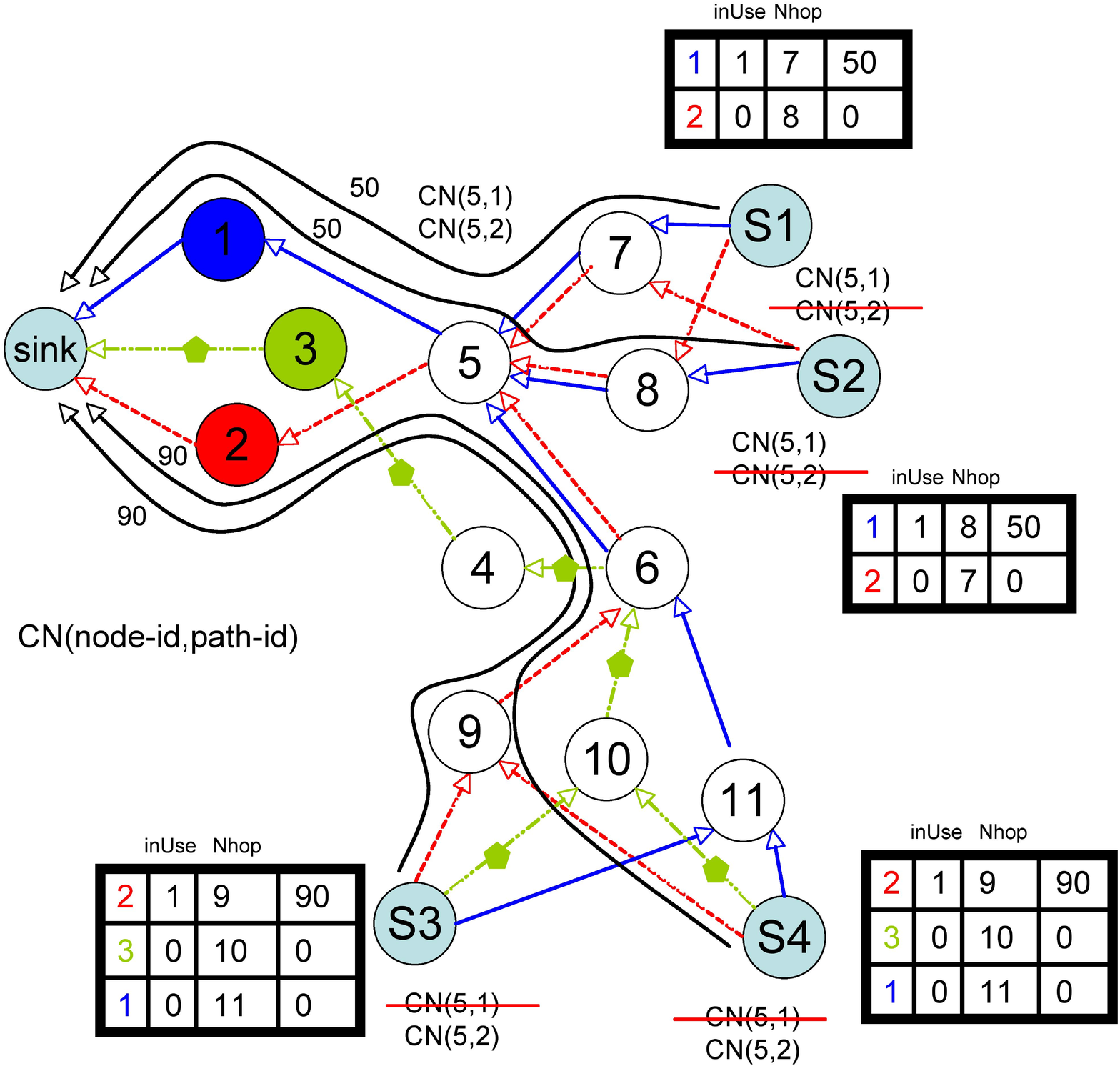,width=.8\linewidth}}
  \caption{Initial configuration, then congestion notification from node 5.}
  \label{MultiPath-2}
\end{figure}

In the scenario depicted by figure \ref{MultiPath-2}, each source
sends a flow of events/messages to the sink. Therefore, according to
the forwarding tables in each source, node 5 sees 4 flows. Assuming
that flows from S1 and S2 are 50kbit/s flows and flows from S3 and S4
are 90kbit/s flows as shown in figure \ref{MultiPath-2}, node 5 has to
relay a total data rate of 280kbit/s.

\begin{figure}
  \centerline{\psfig{figure=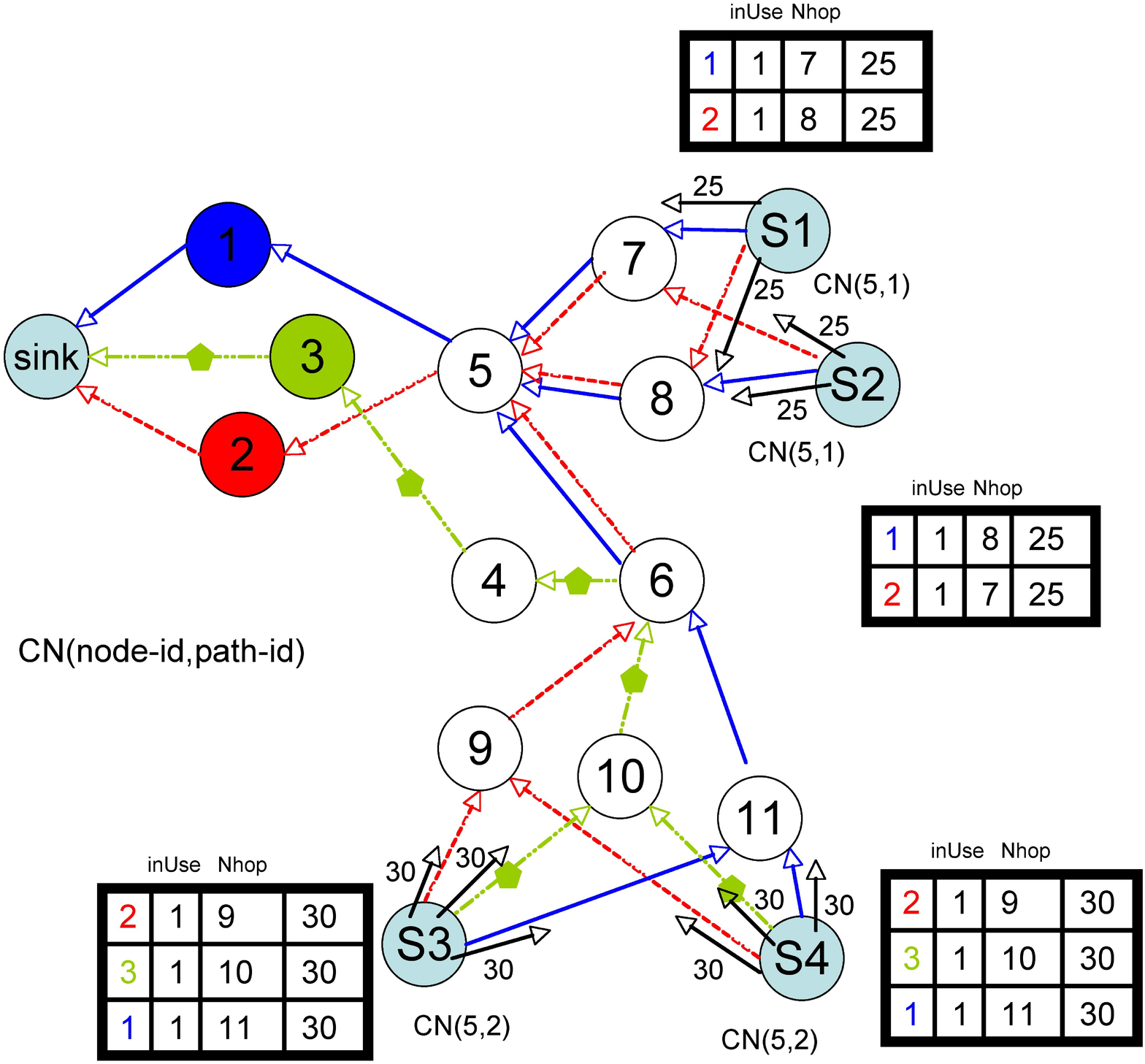,width=.8\linewidth}}
  \caption{After congestion notification from node 5.}
  \label{MultiPath-3}
\end{figure}

If we assume that such a data rate triggers 2 CN messages, CN(5,1) and
CN(5,2), from node 5, sources S1 to S4 will receive them by means of
the intermediate routing nodes. Upon reception of CN messages, each
source will determine which CN message, if any, announces a congestion
on an active path in its local forwarding table. For each active path
$a$, $S_i$ will load-balance its current traffic on $a$ on all the
available paths.  In the scenario of figure \ref{MultiPath-2}, S1 and
S2 will use path id 2 in addition to path id 1 on reception of
CN(5,1), ignoring CN(5,2), and will send on each of these paths
$50/2=25$kbit/s of data. S3 and S4 will use 3 paths on reception of
CN(5,2), and ignoring CN(5,1), by sending $90/3=30$kbit/s of data on
each of those. Finally, we will end up with the data rate repartition
shown in table \ref{repartition}. Node 5 sees a total data rate of
$25+25+25+25+30+30+30+30=220$kbit/s instead of 280kbit/s. Nodes 3, 4 and 10 (on path id 3) relay $30+30=60$kbit/s from sources S3 and S4
(figure \ref{MultiPath-3}). At this stage, with this scenario, mode 3 gives the same repartition than mode 2.  

\begin{table}[H]
\begin{center}
\begin{tabular}{|l|c|c|c|c|c|} \hline \hline
path id   & S1 & S2 & S3 & S4 & total \\ \hline \hline
path id 1 & 25 & 25 & 30 & 30 & 110 \\ \hline
path id 2 & 25 & 25 & 30 & 30 & 110 \\ \hline
path id 3 &    &    & 30 & 30 & 60 \\ \hline 
\hline
total     & 50 & 50 & 90 & 90 & 280 \\ \hline
\hline
\end{tabular}
\end{center}
\caption{Rate repartition after processing CN(5,1) and CN(5,2).}
\label{repartition}
\end{table}

Now, in figure \ref{MultiPath-3}, let us continue to assume that for some
reason node 2 becomes congested (not shown in the figure). At this
point, node 2 is relaying $30+30+25+25=110$kbit/s from 4 flows. In
this case, S1 to S4 will receive a CN(2,2) that will trigger a new
rate repartition. S1 and S2 will then load-balance uniformly their
\underline{current traffic on path id 2} (25 for each from previous
steps) on the 2 available paths. S3 and S4 will also load-balance
uniformly their \underline{current traffic on path id 2} (30 for each
from previous steps) on the 3 available paths. We have now the
repartition illustrated in table \ref{repartition-2}. Node 5 that
previously sent a CN message now have a total data rate of
$25+12.5+25+12.5+12.5+12.5+40+40+10+10=200$kbit/s instead of 220kbit/s
without issuing any CN message. We see that path id 3 now carries a
total traffic of $40+40=80$kbit/s instead of 60kbit/s.

\begin{table}[H]
\begin{center}
\begin{tabular}{|l|c|c|c|c|c|} \hline \hline
path id   & S1 & S2 & S3 & S4 & total \\ \hline \hline
path id 1 & 25+12.5 & 25+12.5 & 40 & 40 & 155 \\ \hline
path id 2 & 12.5 & 12.5 & 10 & 10 & 45 \\ \hline
path id 3 &    &    & 40 & 40 & 80 \\ \hline
total     & 50 & 50 & 90 & 90 & 280 \\ \hline
\hline
\end{tabular}
\end{center}
\caption{Rate repartition after processing CN(2,2).}
\label{repartition-2}
\end{table}

\vspace{-0.5cm}
\section{Simulation Results}\label{simulation}

The routing protocol and the different load repartition strategies
were implemented with TOSSIM, the bit level simulator for TinyOS
platform. We considered a square sensor field of size $1000 \times
1000 m^2$ where a given number of static sensor nodes ranging from 50
to 250 with a step of 50 are randomly deployed. Each node has a
maximum radio range. All the sensors have same processing
capability. We adopted the energy model of \cite{EnergyModel} for
transmission. The energy dissipation due to processing was neglected
in our simulations. The sink is located at the upper right corner
(coordinates 1000,1000) and an event occurrence is simulated at the
opposite quarter of the field. Every video sensor located close enough
to the event will start sensing and transmitting information towards
the sink. Experiments were performed and averaged over 100 simulations
with different randomly generated topologies (with radio range of
400m) and initial energies at the sensor nodes which are generated
following a uniform distribution between $0$ and $0.4$ Joules.

\begin{figure}
  \centerline{
    \includegraphics[width=.95\linewidth,angle=0]{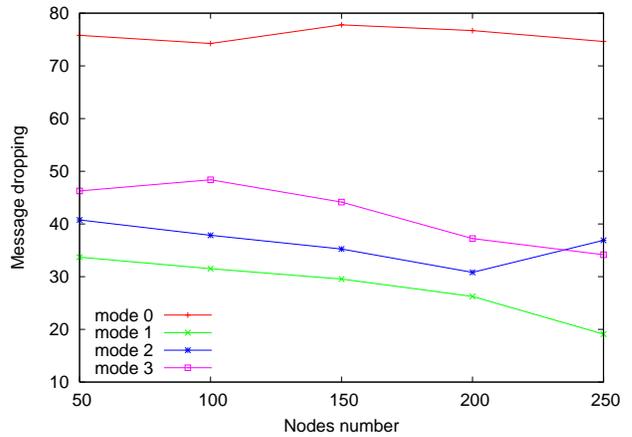}
  }
  \caption{Message dropping rate at sensor queues}
  \label{dropFifo}
\end{figure}

Figure \ref{dropFifo} shows the mean drop rate at the sensor queues as
a function of the number of sensors for the various load repartition
modes. In mode 0, the different sources transmit data with a fixed
rate using only one path without any congestion control. Intermediate
nodes, when overloaded, drop packets and hence the number of dropped
packets is the largest compared to the other modes. Mode 1 gives the
best performances with a dropping rate not more than 35\%. This is due
to the fact that the sources distribute their flows on all available
paths from the beginning hence reducing the probability of overloaded
queues. The other modes appears to have similar performances but less
than mode 1. This is due to the fact that a source sends data on an
other path only when it receives a CN. Meanwhile some packets can be
dropped. However, we see that mode 3 that tries to balance the load of
a congested path on the other paths does not succeed in reducing the
drop rate when compared to a simpler approach such as mode 2, at least
for small network size.

\begin{figure}
  \centerline{
    \includegraphics[width=.95\linewidth,angle=0]{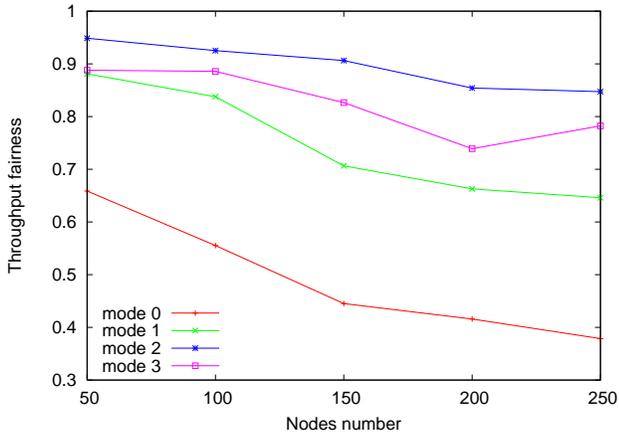}
  }
  \caption{Rate fairness among sources}
  \label{fairness}
\end{figure}

We also looked at the fairness among the sources in term of transmission
rate when performing congestion control. The following commonly used formula:
\begin{eqnarray}
\frac{(\sum_{i=1}^{N_s} r_i)^2}{N_s \sum_{i=1}^{N_s} r_i^2}
\label{fairnessMetric}
\end{eqnarray}
gives a fairness metric where $r_i$ is the success rate achieved by
source $i$ and $N_s$ is the number of sources. Figure \ref{fairness}
shows the achieved fairness among the sources for the different modes
as a function of the number of nodes. It appears that when using only
one path per source (mode 0), fairness among sources is the
worst. This is due to the fact that every source sends without any
coordination since there is no congestion control. When distributing
the flows on all available paths from the beginning (mode 1) without
assessing the congestion situation, we eliminate any coordination
between the sources and fairness among them is difficult to
achieve. In modes 2 and 3, a form of implicit coordination is created
among the sources since a congestion control mechanism using CN is
carried out. We see that in mode 2 for example we achieve a fairness
of more than 80\% even for a large number of nodes.

\begin{figure}
  \centerline{
    \includegraphics[width=.95\linewidth,angle=0]{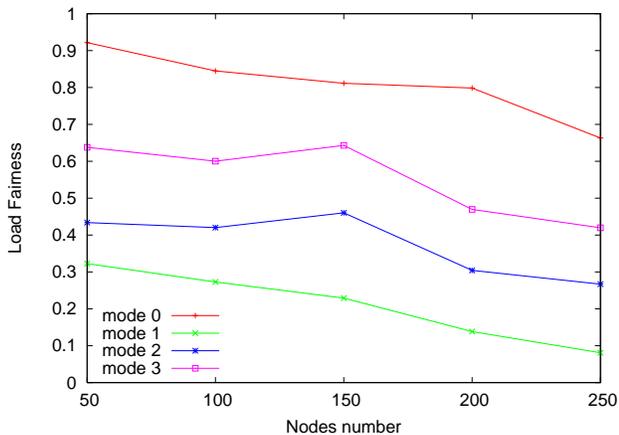}
  }
  \caption{Load fairness among active sensors}
  \label{equiteCharge}
\end{figure}

We also evaluated the load distribution among active sensors
(i.e. those taking part in data forwarding). We used the same fairness
metric but replaced the transmission rate by the amount of processed
data at a given node. Figure \ref{equiteCharge} draws the load
fairness among active sensors for the different modes. Mode 0 achieves
the best fairness since there is only one path and the load fairness
is computed only for sensors belonging to this unique
path. Distributing the flows on all the paths from the beginning (mode
1) does not appear to be interesting from a load distribution
perspective. Mode 3 appears to have the best load distribution since
we take into consideration the load on a per-path basis and then
adjust accordingly the data rate on each active path.

\begin{figure}
  \centerline{
    \includegraphics[width=.95\linewidth,angle=0]{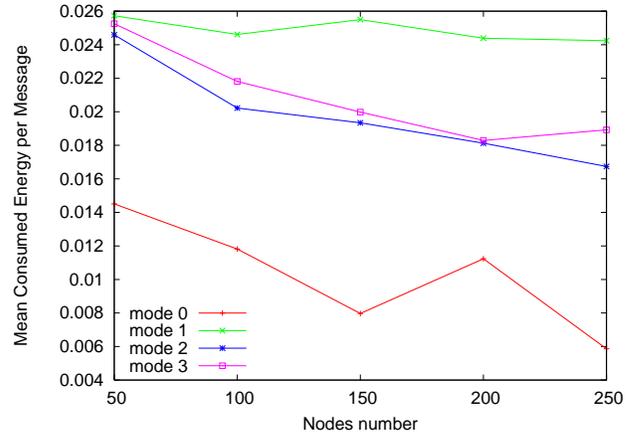}
  }
  \caption{Mean consumed energy per received packet}
  \label{EperMsg}
\end{figure}

Finally, we looked to assess the energy requirements of the different
modes. Figure \ref{EperMsg} shows the amount of consumed energy per
correctly received packet at the sink. Naturally, mode 0 consumes less
energy since, only one path is used and a small amount of data is
received by the sink. Mode 1 consumes more energy per received packet
than the other modes where load repartition mitigates congestion.

\section{Conclusion}\label{conclusion}

In this paper, we investigated the use of load repartition for congestion control of video flows in a
wireless sensor network with multipath support. The motivation of this
study is to maintain the video quality unchanged by splitting a video
flow on multiple path instead of decreasing the transmission
rate. Various load repartition strategies are therefore proposed and
evaluated. The preliminary results show that load repartition does
improve congestion control by reducing the packet drop
probability. Regarding fairness, which is a key factor in congestion
control, the preliminary results show that even simple load
repartition strategies can have a very high impact on
performances. However, depending on the targeted video applications on
the sensor network, one may choose to prefer either rate fairness among
sources or load fairness among active sensors. More importantly, it has
been shown that distributing the traffic on all the available path
from the beginning is not efficient in term of energy nor in term of
fairness.

\bibliographystyle{latex8}
\bibliography{wsn}

\end{document}